\documentclass[12pt]{article}
\usepackage{amsmath,amssymb} 
\usepackage{bm}
\usepackage{subcaption}

\usepackage{amsfonts}
\usepackage{enumerate}
\usepackage{ytableau}
\usepackage{hyperref}
\usepackage{cite}
\usepackage{bm}
\usepackage{tikz}
\usepackage{blkarray}  
\begin{document}
\def\bal#1\eal{\begin{align}#1\end{align}}
	
\def\im{\text{i}}
\def\eqa{\begin{eqnarray}}
\def\eqae{\end{eqnarray}}
\def\be{\begin{equation}}
\def\ee{\end{equation}}
\def\bea{\begin{eqnarray}}
\def\eea{\end{eqnarray}}
\def\ba{\begin{array}}
\def\ea{\end{array}}
\def\bd{\begin{displaymath}}
\def\ed{\end{displaymath}}
\def\eg{{\it e.g.~}}
\def\ie{{\it i.e.~}}
\def\Tr{{\rm Tr}}
\def\tr{{\rm tr}}
\def\>{\rangle}
\def\<{\langle}
\def\a{\alpha}
\def\b{\beta}
\def\c{\chi}
\def\del{\delta}
\def\e{\epsilon}
\def\f{\phi}
\def\vf{\varphi}
\def\tvf{\tilde{\varphi}}
\def\g{\gamma}
\def\h{\eta}
\def\j{\psi}
\def\k{\kappa}
\def\l{\lambda}
\def\m{\mu}
\def\n{\nu}
\def\w{\omega}
\def\p{\pi}
\def\q{\theta}
\def\r{\rho}
\def\s{\sigma}
\def\t{\tau}
\def\u{\upsilon}
\def\x{\xi}
\def\z{\zeta}
\def\D{\Delta}
\def\F{\Phi}
\def\G{\Gamma}
\def\J{\Psi}
\def\L{\Lambda}
\def\W{\Omega}
\def\P{\Pi}
\def\Q{\Theta}
\def\S{\Sigma}
\def\U{\Upsilon}
\def\X{\Xi}
\def\nab{\nabla}
\def\pa{\partial}
\newcommand{\lra}{\leftrightarrow}

\newcommand{\bc}{{\mathbb{C}}}
\newcommand{\br}{{\mathbb{R}}}
\newcommand{\bz}{{\mathbb{Z}}}
\newcommand{\bp}{{\mathbb{P}}}

\def\({\left(}
\def\){\right)}
\def\nn{\nonumber \\}

\newcommand{\red}{\textcolor[RGB]{255,0,0}}
\newcommand{\blue}{\textcolor[RGB]{0,0,255}}
\newcommand{\green}{\textcolor[RGB]{0,255,0}}
\newcommand{\cyan}{\textcolor[RGB]{0,255,255}}
\newcommand{\magenta}{\textcolor[RGB]{255,0,255}}
\newcommand{\yellow}{\textcolor[RGB]{255,255,0}}
\newcommand{\sky}{\textcolor[RGB]{135, 206, 235}}
\newcommand{\orange}{\textcolor[RGB]{255, 127, 0}}
\def\d{\operatorname{d}}
\def\ttbar{T$\overline{\text{T}}$ }
\def\Renyi{R$\acute{\text{e}}$nyi }
\def\Poincare{Poincar$\acute{\text{e}}$ }
\def\Banados{Ba$\tilde{\text{n}}$ados }

\title{\textbf{Modular transformations of on-shell actions of (root-)\ttbar deformed holographic CFTs}}
\vspace{14mm}
\author{Jia Tian$^{1,2}$\footnote{wukongjiaozi@ucas.ac.cn}, Tengzhou Lai$^{3}$\footnote{laitengzhou20@mails.ucas.ac.cn} and Farzad Omidi$^{2}$\footnote{omidi@ucas.ac.cn}}
\date{}
\maketitle

\begin{center}
	{\it $^1$State Key Laboratory of Quantum Optics and Quantum Optics Devices, Institute of Theoretical Physics, Shanxi University, Taiyuan 030006, P.~R.~China\\
		\vspace{2mm}
		$^2$Kavli Institute for Theoretical Sciences (KITS),\\
		University of Chinese Academy of Science, 100190 Beijing, P.~R.~China \\
		\vspace{2mm}
		$^3$School of Physical Sciences, University of Chinese Academy of Sciences, Zhongguancun East Road 80, Beijing 100190, P.~R.~China
	}
	\vspace{10mm}
\end{center}
\begin{abstract}
In this study, we examine the modular transformations of the (root-) \ttbar deformed torus partition function of a two-dimensional CFT (with a gravitational anomaly) from the holographic perspective by computing the on-shell actions of various saddle solutions of the dual gravity theories. 
\end{abstract}

\newpage

\tableofcontents
\section{Introduction}
In the past two decades, a very interesting irrelevant deformation of two-dimensional conformal field theories (CFTs), known as {\it \ttbar deformation}, was proposed \cite{Zamolodchikov:2004ce,Smirnov:2016lqw,Cavaglia:2016oda}. One of the most intriguing properties of \ttbar deformed CFTs is that one can exactly calculate some quantities including energy spectrum \cite{Zamolodchikov:2004ce,Smirnov:2016lqw,Cavaglia:2016oda} and S-Matrix \cite{Smirnov:2016lqw,Dubovsky:2017cnj}.\footnote{Moreover, the closed form of the Lagrangian of the \ttbar deformed theory can be obtained for some models, e.g. \cite{Bonelli:2018kik,Chen:2020jdi}.}
In recent years, various features of the \ttbar deformed CFTs were investigated such as Lax connections\cite{Conti:2018jho,Chen:2021aid}, S-Matrix \cite{Dubovsky:2017cnj,Dey:2021jyl}, correlation functions \cite{Guica:2019vnb,Chakraborty:2023wel,Kraus:2018xrn,Aharony:2018vux,Cardy:2019qao,He:2019vzf,He:2019ahx,He:2020udl,Cui:2023jrb,He:2023kgq,He:2023hoj,He:2022jyt,He:2020qcs,Aharony:2023dod,Castro-Alvaredo:2023wmw,Li:2020pwa,Hirano:2020ppu,He:2023knl}, partition function\cite{Cardy:2018sdv,Datta:2018thy,Apolo:2023aho,Aharony:2018bad,He:2020cxp,He:2024pbp,Datta:2021kha,Cardy:2022mhn,Bhattacharyya:2023gvg,Kraus:2022mnu}, entanglement entropy \cite{Donnelly:2018bef,Chen:2018eqk,Park:2018snf,Banerjee:2019ewu,Jeong:2019ylz,Grieninger:2019zts,Murdia:2019fax,Donnelly:2019pie,Jia:2023ttbar,Allameh:2021moy,FarajiAstaneh:2022qck,He:2023xnb,Ebrahim:2023ush,Jeong:2022jmp,Castro-Alvaredo:2023jbg,Apolo:2023vnm,Grieninger:2023knz,Jiang:2023ffu,Jiang:2023loq,He:2023obo,Basu:2024bal}, pseudo entanglement entropy \cite{Chen:2023eic,He:2023wko}, entanglement negativity \cite{Basu:2023bov}, mutual information \cite{Asrat:2020uib,Khoeini-Moghaddam:2020ymm,Ebrahim:2023ush}, entanglement wedge cross section \cite{Asrat:2020uib,Khoeini-Moghaddam:2020ymm,Basu:2023aqz,Basu:2024bal} and Action-Complexity \cite{Akhavan:2018wla,Alishahiha:2019lng,Hashemi:2019xeq}.
Furthermore, theses deformations were extended to quantum field theories in zero dimension \cite{Das:2022uhj}, one dimension \cite{Gross:2019ach,Gross:2019uxi,He:2021dhr}, higher dimensions \cite{Hartman:2018tkw,Bonelli:2018kik,Taylor:2018xcy,Hou:2022csf,Morone:2024ffm,Ebert:2024fpc}, to non-relativistic quantum field theories \cite{Cardy:2018jho,Alishahiha:2019lng,Chen:2020jdi,Khoeini-Moghaddam:2020ymm} as well as to lattice theories \cite{Jiang:2023rxa}. See also \cite{Jiang:2019epa,Monica} which are very nice and comprehensive reviews on the recent developments. 
\\On the other hand, there is another interesting deformation of CFTs dubbed {\it Root-\ttbar deformation} which is marginal and commutes with the \ttbar deformation \cite{Rodriguez:2021tcz,Tempo:2022ndz,Babaei-Aghbolagh:2022uij,Ferko:2022cix,Borsato:2022tmu,Babaei-Aghbolagh:2022leo,Ebert:2024zwv,Ebert:2023tih}.  Moreover, similar to the \ttbar deformation, the deformation operator does not change under the flow \cite{Ebert:2023tih}. However, it preserves the conformal invariance in the sense that the trace of the stress energy tensor remains zero under the flow \cite{Borsato:2022tmu,Ebert:2023tih}.
\\In the context of the AdS/CFT correspondence \cite{Maldacena:1997re} , there are two different proposals for the holographic dual of \ttbar deformed CFTs:
\footnote{The holographic dual of root-\ttbar deformed CFTs was studied in \cite{Ebert:2023tih}.} 
cut-off  proposal \cite{McGough:2016lol,Kraus:2018xrn} (or glue-on proposal \cite{Apolo:2023vnm,Apolo:2023ckr}) and mixed-boundary-condition proposal \cite{Guica:2019nzm}. The cut-off proposal is easier to  implement, but it has several limitations\cite{Guica:2019nzm}. On the other hand, the mixed-boundary-condition proposal is applicable to all holographic theories, regardless of the presence of matters, and the holographic dictionary can be summarized as follows: \cite{Monica} \footnote{The notation will become clear in a moment.}
\bea\label{dictionary}
Z_{\text{\ttbar,CFT}}[\gamma^{[\mu]}_{\alpha\beta}]=Z_{\text{grav}}\left[g_{\alpha\beta}^{(0)}+\frac{\mu}{16\pi G_N}g_{\alpha\beta}^{(2)}+\frac{\mu^2}{(16\pi G_N)^2}g_{\alpha\beta}^{(4)}=\gamma^{[\mu]}_{\alpha\beta}\right],
\eea 
where $\mu$ is the deformation parameter. The key observation of us is that in the dictionary \eqref{dictionary} we should add a proper boundary term to the gravity action {(See eq. \eqref{bdy})}
\footnote{Perhaps this is already hidden written in \cite{Guica:2019nzm}
and the presence of this {term} is stressed by one of the authors in \cite{Jia:2023ttbar}.}. 
The presence of this boundary term is natural considering that the double trace deformation not only shifts the source but also shifts the generating function \cite{Klebanov:1999tb,Witten:2001ua}. In this paper, we compute the on-shell actions of the \ttbar deformed theory defined on a torus. The aim is to derive and understand the modular properties of the \ttbar deformed torus partition function from the holographic perspective. Interestingly, the new boundary term proves to be crucial for the on-shell action to exhibit the appropriate modular property and other desirable characteristics. 
{It is also worth noting that in the cut-off proposal of the \ttbar deformation, the presence of a similar term has been suggested \cite{Caputa:2020lpa}. However, in the cut-off proposal, the boundary term is introduced at the cut-off surface instead of the AdS boundary. }

A two-dimensional CFT defined on a torus (and other Riemann surfaces) is subject to an important set of consistent conditions known as modular invariance. This requires that the partition function or correlation functions should remain the same, regardless of how the torus is represented. Modular invariance is a powerful constraint, leading to significant consequences such as the Hawking-Page transition \cite{Hawking:1982dh} and the Cardy formula \cite{Cardy:1986ie}. It has been discovered in \cite{Datta:2018thy} {and} also in \cite{Apolo:2023aho} that the deformation parameter should also transform under the modular transformation which means that this kind of transformation can relate the partition functions of two different quantum field theories along the \ttbar flow. We find that this property can be easily understood by observing that the deformation parameter $\mu$ has the same dimension of the radial coordinate  and undergoes rescaling under a Weyl transformation. Meanwhile, we derive the thermodynamic first law and Smarr relations for the deformed BTZ black hole. Furthermore, we show that the modular property remains preserved even in the CFT with a gravitational anomaly\cite{Alvarez-Gaume:1983ihn}, where the holomorphic sector and anti-holomorphic sector have different central charges\cite{Kraus:2005zm}, which is dual to a topologically massive gravity (TMG) theory \cite{Deser:1982vy, Deser:1981wh}. As a contrast, we also consider a marginal deformation, the root-\ttbar deformation \cite{Ferko:2022cix,Borsato:2022tmu} in the mixed-boundary-condition proposal \cite{Ebert:2023tih}, for which the deformation parameter is dimensionless and we find indeed it does not transform under modular transformation.

\section{Mixed-boundary-condition proposal and the deformed on-shell action}
\label{section:reviewttbar}
\renewcommand{\theequation}{2.\arabic{equation}}
\setcounter{equation}{0}
In this section, we briefly review the mixed-boundary-condition proposal of the \ttbar deformation with a special emphasis on the deformed on-shell action.
The \ttbar deformation of 2d field theory with action $S_{\text{CFT}}$ is defined by the flow equation
\bea
\frac{dS^{[\mu]}_{\text{CFT}}}{d\mu}=\int d^2 x \sqrt{\gamma}\, \text{T}\bar{\text{T}}^{[\mu]},
\eea 
where the  double trace operator $\text{T}\bar{\text{T}}^{[\mu]}$ is defined as
\bea
\text{T}\bar{\text{T}}^{[\mu]}=\frac{1}{8}\(T_{\alpha\beta}T^{\alpha\beta}-(T^\alpha_\alpha)^2\),
\eea 
and it is evaluated in the deformed theory with deformation parameter $\mu$.
 In the linear order, the deformation is  a special double trace deformation
\bea \label{fieldaction}
S_{CFT}^{[\mu]}=S_{CFT}+\mu \int d^2 x \sqrt{\gamma}\, \text{\ttbar}+\mathcal{O}(\mu^2).
\eea 
The  significance of this double trace deformation is that the operator \ttbar depends both on the boundary metric and the energy-stress tensor. Consequently, both of them flow under the deformation according to \cite{Guica:2019nzm}:
\bea 
&&\gamma_{\alpha\beta}^{[\mu]}=\gamma_{\alpha\beta}^{[0]}+\frac{1}{2}\mu \hat{T}_{\alpha\beta}^{[0]}+\frac{1}{16}\mu^2\hat{T}_{\alpha\rho}^{[0]}\hat{T}_{\sigma\beta}^{[0]}{\gamma^{[0]}}^{\rho\sigma},\label{metricflowm}\\
&&\hat{T}^{[\mu]}_{\alpha\beta}=\hat{T}^{[0]}_{\alpha\beta}+\frac{1}{4}\mu\hat{T}_{\alpha\rho}^{[0]}\hat{T}_{\sigma\beta}^{[0]}{\gamma^{[0]}}^{\rho\sigma},\label{energyflowm}
\eea 
where $\hat{T}_{\alpha\beta}=T_{\alpha\beta}-\gamma_{\alpha\beta}T$, see Appendix \ref{append:ttbar} for {more} details. The initial data $(\gamma_{\alpha\beta}^{[0]},\hat{T}^{[0]}_{\alpha\beta})$ defines the seed CFT which generally should not be considered as the undeformed CFT because from the field theory point of view the \ttbar deformation only deforms the theory but not the spacetime background\footnote{Do not be confused with the dynamical coordinate transformation interpretation \cite{Conti:2018tca,Caputa:2020lpa} of the \ttbar deformation.}. The seed CFT is dual to an asymptotic AdS$_3$ solution and its metric can be written in the Fefferman-Graham gauge {as follows} \cite{Skenderis:1999nb}
\bea 
ds^2=g_{\alpha\beta}(z,x^\alpha)dx^\alpha dx^\beta+\frac{dz^2}{z^2},\quad g_{\alpha\beta}(z,x^\alpha)=\frac{g_{\alpha\beta}^{(0)}}{z^2}+g_{\alpha\beta}^{(2)}+z^2 g_{\alpha\beta}^{(4)},\quad g_{\alpha\beta}^{(0)}=\gamma_{\alpha\beta}^{[0]},
\eea 
where $g^{(2)}$ corresponds to the expectation value of the seed CFT operator
\bea \label{energymap}
\hat{T}^{[0]}_{\alpha\beta}=\frac{1}{8\pi G_N}g_{\alpha\beta}^{(2)}.
\eea 
The solution \eqref{metricflowm} implies that the deformed boundary metric can be interpreted as the induced metric at the cut-off surface $z=z_c\equiv\sqrt{\rho_c}$:
\bea 
\gamma_{\alpha\beta}^{[\mu]}=g_{\alpha\beta}^{(0)}+\frac{\mu}{16\pi G_N}g_{\alpha\beta}^{(2)}+\frac{\mu^2}{(16\pi G_N)^2}g_{\alpha\beta}^{(4)}=\rho_c g_{\alpha\beta}(z_c=\sqrt{\rho_c}),\quad \rho_c=\frac{\mu}{16\pi G_N }. \label{gammamu}
\eea 
{In the following,} we will interchangeably use $\rho_c$ and $\mu$ as the deformation parameters. 
\\As pointed out in \cite{Caputa:2020lpa,Jia:2023ttbar}, the proper on-shell action of the \ttbar deformed theory should be 
\bea 
I_{\text{on-shell}}^{[\mu]}&=&I_{\text{Euclidean}}^{[0]}\(\gamma^{[\mu]}_{\alpha\beta}=g_{\alpha\beta}^{(0)}+\frac{\mu}{16\pi G_N}g_{\alpha\beta}^{(2)}+\frac{\mu^2}{(16\pi G_N)^2}g_{\alpha\beta}^{(4)}\)-\mu\int\sqrt{\gamma^{[\mu]}}\text{T}\bar{\text{T}}^{[\mu]} \nonumber\\
&\equiv &I_{\text{bulk}}^{[\mu]}+I_{\text{bdy}}^{[\mu]}.\label{onshell}
\eea 
Using \eqref{metricflowm} and \eqref{energyflowm}, one can show that $\sqrt{\gamma}\text{\ttbar}$ is invariant under the flow, so the boundary term in \eqref{onshell} can also be written as
\bea 
-\mu\int\sqrt{\gamma^{[\mu]}}\text{T}\bar{\text{T}}^{[\mu]}=-\mu \int \sqrt{g^{[0]}}\text{T}\bar{\text{T}}^{[0]}. \label{bdy}
\eea  
When the seed metric $g_{\alpha\beta}^{(0)}$ is flat, the bulk solution in the FG gauge is characterized by two arbitrary functions as
\be  \label{FG}
ds^2=\frac{dz^2}{z^2}+\frac{1}{z^2} \(dv+\bar{L}_0 (\bar{v})z^2 d\bar{v} \) \(d\bar{v}+L_0(v) z^2 dv \).
\ee 
In this case, one can show that the deformed metric $\gamma_{\alpha\beta}^{[\mu]}$ is also flat thus \eqref{gammamu} is related to $g_{\alpha\beta}^{(0)}$ via the {following} coordinate transformations
\bea 
&&dVd\bar{V}=\(dv+\bar{L}_0(\bar{v})\rho_c d\bar{v}\)\(d\bar{v}+{L}_0({v})\rho_c d{v}\),\\
&&V=v+\rho_c\int \bar{{L}}_0({\bar{v}})d\bar{v},\quad \bar{V}=\bar{v}+\rho_c\int L_0(v)dv. \label{trans}
\eea 
Substituting the inverse of \eqref{trans} into \eqref{FG} leads to the metric of the dual geometry of the \ttbar deformed CFT
\bea \label{demetric}
ds_\mu^2&=&\frac{dz^2}{z^2}+\frac{1}{z^2(1-\rho_c^2 L_0\bar{L}_0)^2}\((1-\rho_cL_0\bar{L}_0z^2)dV+\bar{L}_0(z^2-\rho_c)d\bar{V}\) \nonumber\\
\qquad &&\((1-\rho_cL_0\bar{L}_0z^2)d\bar{V}+{L}_0(z^2-\rho_c)d{V}\), 
\eea 
where $L_0$ and $\bar{L}_0$ should be understood as functions of $V$ and $\bar{V}$. {Moreover,} the deformed stress tensor is given by \eqref{energyflowm}:
\be \label{destress}
T_{\alpha\beta}^{[\mu]}dx^\alpha dx^\beta=\frac{L_0dV^2+\bar{L}_0d\bar{V}^2+2\rho_c L_0\bar{L}_0dVd\bar{V}}{8\pi G_N (1-\rho_c^2L_0\bar{L}_0)},
\ee 
and the deformed on-shell action is equal to \cite{Jia:2023ttbar}
\be \label{deformedonshell}
I_{\text{on-shell}}^{[\mu]}=-\frac{c}{6\pi}\int \sqrt{g^{[0]}}\(\sqrt{L_0\bar{L}_0}+\rho_c L_0\bar{L}_0\),
\ee 
where $c=3/(2G_N)$ \cite{Brown:1986nw}.
Since we are interested in the modular property of the torus partition function, below we will assume that the boundary manifold is a flat torus with two generic complex periods $\bm{\w}=\omega_1+\im \omega_2$ and $\bm{\beta}=\beta_1+\im \beta_2$. Therefore, the possible bulk solutions are global AdS, BTZ black hole and other $SL(2,R)$ black holes. Because AdS$_3$ gravity does not have local degrees of freedom, these solutions are all diffeomorphic equivalent but distinguished by the periodicity.
A torus is specified by the modular parameter  defined by the ratio of the two complex periods: $\bm{\tau}=\frac{\bm{\beta}}{\bm{\omega}}$. It is well known that the undeformed on-shell actions of these solutions are related to each other via the $SL(2,\mathbb{Z})$ modular transformation. This modular property can be traced back to the diffeomorphic equivalence of bulk solutions. However, it may be broken due to the non-commutativity of the \ttbar deformation with the diffeomorphism. Nevertheless, \cite{Datta:2018thy} shows that the \ttbar deformed torus partition function exhibits a modified modular invariance:
\be 
Z\(\bm{\tau}\Big|\rho_c\)=Z\( \frac{a\bm{\tau}+b}{c\bm{\tau}+d}\Big|\frac{\rho_c}{|c\bm{\tau}+d|^2}\), \label{moz}
\ee   
by directly calculating the torus partition function of a \ttbar deformed 2d CFT.
In this work, we aim to understand this modular property holographically by computing the deformed on-shell actions of the dual gravity theory. Before turning on the \ttbar deformation, we will construct the global AdS and black hole solutions with two complex periods $\bm{\w}$ and $\bm{\beta}$ as a prerequisite.
\\We want to emphasize that during the \ttbar deformation process, three distinct CFTs are involved: the seed CFT defined on the seed torus with metric $ds^2=dvd\bar{v}$, the deformed CFT defined on the deformed torus with metric $ds^2=dVd\bar{V}$, and the undeformed CFT defined on the undeformed torus with metric $d\tilde{s}^2=d\tilde{v}d\bar{\tilde{v}}$. The explicit relationships among these three CFTs will be discussed in each of the subsequent examples.
\section{General global AdS}
\renewcommand{\theequation}{3.\arabic{equation}}
\setcounter{equation}{0}
Let us start from the standard global AdS solution with\footnote{In this work, we set the AdS radius to be $\ell=1$.}
\be 
L_0=\bar{L}_0=-\frac{1}{4},\quad \bm{\w}=2\pi,\quad \bm{\beta}=2\pi \bm{\tau},
\ee 
whose metric is given by
\be  \label{eq:sglobal}
ds^2=\frac{dz^2}{z^2}+\frac{(z^2-4)^2d\phi^2}{16z^2}+\frac{(z^2+4)^2 dt^2}{16 z^2},\quad \phi+\im t\equiv v\, ,
\ee
which can be transformed into a more familiar form
 by introducing the radial coordinate $ 
r=|\frac{(z^2-4)}{4z}|.
$
The modular parameter $\bm{\tau}$ can be chosen freely and later on we will use this freedom to construct other $SL(2,\mathbb{Z})$ black holes following \cite{Dijkgraaf:2000fq}. A global AdS with a general $\bm{\omega}$ can be obtained by first rescaling the coordinate $v$ as 
\be \label{rescalingc}
 v \rightarrow \frac{2\pi}{\bm{\w}}v,
\ee 
 and then adjusting the radial coordinate $z$ accordingly such that the metric is in the standard form \eqref{FG}. We find that 
 \be \label{gglobalL}
{L}_0=-\frac{\pi^2}{\bm{\w}^2},\quad \bar{{L}}_0=-\frac{\pi^2}{\bar{\bm{\w}}^2},
\ee 
and the resulting metric is
\be 
ds^2=\frac{dz^2}{{z}^2}+g_{{t}{t}}d{t}^2+\frac{(1+{L}_0 {z}^2)(1+\bar{{L}}_0{z}^2)}{{z}^2}(d{\phi}+\Omega d{t})^2,
\ee 
where
\be 
g_{{t}{t}}=\frac{(1-{z}^4 {L}_0\bar{{L}}_0)^2}{{z}^2(1+{L}_0{z}^2)(1+\bar{{L}}_0 {z}^2)},\quad\Omega=\frac{\im(\bar{{L}}_0-{L}_0){z}^2}{(1+{L}_0 {z}^2)(1+\bar{{L}}_0{z}^2)}.
\ee 
Note that the metric will have no horizons  when ${L}_0=\bar{{L}}_0$ which means $\bm{\w}=\bar{\bm{\w}}$.
\subsection{\ttbar deformation} 
According to the dictionary \eqref{trans}, the \ttbar deformed theory is also defined on a torus with a flat metric $ds^2=dVd\bar{V}$, which is related to the seed metric via
\bea \label{relation1}
&&V=v+\bar{L}_0 \rho_c \bar{v},\quad \bar{V}=\bar{v}+L_0 \rho_c v,\\
&&v=\frac{V-\rho_c\bar{L}_0 \bar{V}}{1-\rho_c^2 L_0\bar{L}_0},\quad \bar{v}=\frac{\bar{V}-\rho_c {L}_0 {V}}{1-\rho_c^2 L_0\bar{L}_0},\label{relation2}
\eea
where $L_0$ and $\bar{L}_0$ are given by \eqref{gglobalL}.
It implies that the two periods of the deformed torus are
\be 
\bm{\w}_\mu=\bm{\w}+\bar{L}_0\rho_c{\bm{\bar{\w}}}=\bm{\omega} \left(1-\frac{\pi^2\rho_c}{\bm{\w}^2} \right),\quad \bm{\beta}_\mu=\bm{\beta}-\frac{\pi^2\rho_c}{\bm{\w}^2} \bm{\bar{\beta}},\label{wmu}
\ee 
and the \ttbar deformed bulk geometry is described by the metric
\bea \label{gdemetric}
ds_\mu^2=\frac{dz^2}{z^2}-\frac{\pi^2}{\bm{\w}^2}(dv^2+d\bar{v}^2)+\frac{(\bm{\w}^4+\pi^4 z^4)dvd\bar{v}}{\bm{\w}^4z^2},
\eea 
with \eqref{relation2}. 
Using \eqref{destress} we find that the deformed holographic mass and momentum are
\bea \label{globalmu}
M_\mu &=&\frac{c\bm{\w}_\mu}{6\pi}\frac{L_0}{1+\rho_c L_0}=-\frac{c\pi}{3(\bm{\w}_\mu+\sqrt{\bm{\w}_\mu^2+4\pi^2\rho_c})},\\
J_{\mu}&=&0,
\eea 
Defining $M_0=M_\mu|_{\mu=0}=-\frac{c\pi}{6\bm{\w}_\mu}$, we find that the deformed energy can be rewritten as
\be \label{mum0}
M_\mu=\frac{2\bm{\w}_\mu}{\mu}\(1-\sqrt{1-\mu \frac{{M}_0}{\bm{\w}_\mu}}\),\quad \mu={16\pi G_N }\rho_c\, ,
\ee 
which is consistent with the field theory calculation of the deformed energy \cite{Zamolodchikov:2004ce,Smirnov:2016lqw,Cavaglia:2016oda}. 
It may be confusing that the undeformed energy $M_0$ depends on $\bm{\w}_\mu$ which depends on the deformation parameter $\rho_c$ implicitly via \eqref{wmu}. However, as we emphasized in last section when we compare the deformed CFT and the undeformed one, we should fix the size of the system. That is why the undeformed energy $M_0$ can be obtained directly by setting $\rho_c=0$ in \eqref{globalmu}. 
\\The deformed on-shell action can be easily worked out by substituting \eqref{gglobalL} into \eqref{deformedonshell} and rewriting $(\bm{w},\bm{\beta})$ in terms of $(\bm{\w}_\mu,\bm{\beta}_\mu$):
\bea \label{eq:onshell1}
I_E^{[\mu]}&=&
-\frac{c\pi A_\mu}{3}\frac{1}{|\bm{\w}_\mu|^2+|\bm{\w}_\mu|\sqrt{4\pi^2\rho_c+|\bm{\w}_\mu|^2}}\\
&=&-\frac{cA_\mu}{6\pi}\frac{\pi^2}{|\bm{\w}_\mu|^2}{+}\rho_c\frac{cA_\mu}{6\pi}\frac{\pi^4}{|\bm{\w}_\mu|^4}+\mathcal{O}(\rho_c^2),\nonumber\\
&=&I_E^{[0]}[\bm{\w}_\mu]+\mu \int d^2 x \sqrt{\gamma}\, \text{\ttbar}+\mathcal{O}(\mu^2), \label{gonshellexpand}
\eea 
where $A_\mu=|\bm{\w}_\mu|^2\text{Im}(\bm{\tau}_\mu)$ is the area of the deformed torus. In particular, the on-shell action equals to the free energy
\be 
I_E^{[\mu]}=\text{Im}(\bm{\beta}_\mu)M_\mu,
\ee 
and the expansion \eqref{gonshellexpand} coincides exactly with the field theory result \eqref{fieldaction}.

\section{The BTZ black hole}
\renewcommand{\theequation}{4.\arabic{equation}}
\setcounter{equation}{0}
A Euclidean rotating BTZ black hole has the standard metric
\be\label{eq:sbtz}
ds^2=\frac{(r^2-R_+^2)(r^2+R_-^2)}{r^2}dt^2+\frac{r^2}{(r^2-R_+^2)(r^2+R_-^2)}dr^2+r^2(d\phi-\frac{R_+R_-}{r^2}dt)^2,
\ee 
where
\bea 
(\phi,t)\sim (\phi+2\pi,t),\quad (\phi,t)\sim (\phi+\theta,t+\beta).
\eea 
Moreover, its two (complex) periods $\beta$ and $\theta$ are related to the inner and outer horizon radii via
\be 
\beta=\frac{2\pi R_+}{R_+^2+R_-^2},\quad \theta=\frac{2\pi R_-}{R_+^2+R_-^2}.
\ee 
To obtain a BTZ black hole solution with two generic complex periods $(\bm{\w}=\w_1+\im \w_2,\bm{\beta}=\beta_1+\im \beta_2)$, we can perform a similar rescaling as \eqref{rescalingc} to the standard BTZ metric \eqref{eq:sbtz}. Transforming the resulting metric into the FG gauge, we find that
\be \label{bhl0}
L_0=-\frac{\pi^2}{\bm{\beta}^2},\quad \bar{L}_0=-\frac{\pi^2}{\bm{\bar{\beta}}^2}.
\ee 
For more details refer to Appendix \ref{generalEM}.
Because the effect of the rescaling is similar to a boost (or more precisely a rotation in the Euclidean signature), the energy and  momentum get transformed accordingly and  take a slightly complicated form  
\bea 
M&=&-\frac{c\pi}{6|\bm{\beta}|^4}\(\omega_1(\beta_1^2-\beta_2^2)+2\omega_2\beta_1\beta_2\),\label{eqm}\\
J&=&-\frac{c\pi}{6|\bm{\beta}|^4}\(\omega_2(\beta_1^2-\beta_2^2)-2\omega_1\beta_1\beta_2\), \label{eqj} 
\eea 
while the black hole entropy is invariant under the boost and manifestly it only depends on the modular parameter
\bea 
S_{BH}=\frac{c\pi}{3}\frac{\text{Im}(\bm{\tau})}{|\bm{\tau}|^2}.
\eea 
The details of  these derivations can be found in Appendix \ref{generalEM}.
It is straightforward to check that the Smarr relation
\be
S_{BH}=2(\beta_2 M+\beta_1 J),
\ee 
and the Cardy formula
\be 
S_{BH}=\sqrt{\frac{c\pi\bm{\w}}{6}{(M+\im J)}{}}+\sqrt{\frac{c\pi\bm{\bar{\w}}}{6}{(M-\im J)}{}},
\ee 
are satisfied. 
\subsection{The \ttbar deformation}
The \ttbar deformed theory is also defined on a torus with metric $ds^2=dVd\bar{V}$ where the deformed coordinates $(V,\bar{V})$ are related to {the} seed metric via \eqref{relation1} and \eqref{relation2}, however now $L_0$ and $\bar{L}_0$ are given by \eqref{bhl0}. It implies that the two complex periods of the deformed torus are
\be 
\bm{\w}_\mu=\bm{\w}-\frac{\pi^2\rho_c}{\bm{\bar{\beta}}^2}\bm{\bar{\w}},\quad \bm{\beta}_\mu=\bm{\beta}(1-\frac{\pi^2\rho_c}{|\bm{\beta}|^2}),\label{bmu}
\ee 
and the \ttbar deformed bulk geometry is described by the corresponding metric \eqref{demetric}.
The relation \eqref{bmu} implies that $\bm{\omega}_\mu$ is complex in general and the geometry describes a general deformed $SL(2,\mathbb{Z})$ black hole which we will consider later on. For BTZ solutions, we choose $\bm{\w}_\mu$ to be real. The energy and momentum of the deformed theory can be found directly by using \eqref{destress} as follows
\bea 
M_\mu &=&
\frac{c\bm{\w}_\mu}{12\pi }\frac{1}{\rho_c}\(1-\frac{|\bm{\beta}_\mu|}{\sqrt{|\bm{\beta}_\mu|^2+4\pi^2\rho_c}}\frac{|\bm{\beta}_\mu|^4+(\bm{\beta}_\mu+\bm{\bar{\beta}}_\mu)^2\pi^2\rho_c}{|\bm{\beta}_\mu|^4}\),\\
J_\mu &=&
\frac{c\pi\bm{\w}_\mu }{12}\frac{\im (\bm{\bar{\beta}}_\mu^2-\bm{\beta}_\mu^2)}{|\bm{\beta}_\mu|^3\sqrt{|\bm{\beta}_\mu|^2+4\pi^2\rho_c}}. 
\eea 
In the limit $\rho_c \rightarrow 0$, they reduce to
\bea  
M_0&=&M_{\mu}\Big|_{\mu=0}=-\frac{c\pi\bm{\w}_\mu }{12}\(\frac{1}{\bm{\beta}_\mu^2}+\frac{1}{\bm{\bar{\beta}}_\mu^2}\),\\
J_0&=&J_{\mu}\Big|_{\mu=0}=-\frac{c\pi \bm{\w}_\mu}{12}\im\(\frac{1}{\bm{\bar{\beta}}_\mu^2}-\frac{1}{\bm{\beta}_\mu^2}\).
\eea 
However, $M_0$ and $J_0$ are not the undeformed energy and momentum. To see this, for example  we set $J_0=J_\mu=0$ (i.e. $\bm{\beta}_{\mu}=-\bm{\bar{\beta}}_{\mu}$), then we find 
\be 
M_\mu=\frac{2\bm{\w}_\mu}{\mu}\(1-\frac{1}{\sqrt{1+\mu \frac{M_0}{\bm{\w}_\mu}}}\),
\ee 
which is different from the field theory expectation \eqref{mum0}.
The mismatch is due to the fact that the \ttbar deformation alters the temperature of the system! 
In fact, the undeformed CFT lives on the torus with the flat metric $d\tilde{s}^2=d\tilde{v}d\bar{\tilde{v}}$ and coordinate $\tilde{v}$ is defined by \cite{Guica:2019nzm}
\bea \label{tildev}
\tilde{v}=\frac{V-\rho_c \bar{L}_0 \bar{V}}{1-\rho_c\bar{L}_0}=\frac{1-\rho_c^2 L_0\bar{L}_0}{1-\rho_c\bar{L}_0}v,\quad \bar{\tilde{v}}=\frac{\bar{V}-\rho_c {L}_0 {V}}{1-\rho_c{L}_0}=\frac{1-\rho_c^2 L_0\bar{L}_0}{1-\rho_c{L}_0}\bar{v},
\eea 
where $L_0$ and $\bar{L}_0$ are given by \eqref{bhl0}.
Therefore, the two complex periods of this undeformed torus are 
\bea 
\bm{\tilde{\w}}=\bm{\w}_\mu,\quad \bm{\tilde{\beta}}=\bm{\beta}_\mu \frac{1+\frac{\rho_c\pi^2}{|{\bm{\beta}}|^2}}{1+\frac{\rho_c\pi^2}{\bm{\bar{\beta}}^2}}=\frac{\left|\bm{\b}_{\mu}\right|^2\sqrt{\left|\bm{\b}_{\mu}\right|^2+4\pi^2\r_c}}{\text{Re}\(\bm{\b}_{\mu}\)\sqrt{\left|\bm{\b}_{\mu}\right|^2+4\pi^2\r_c}-\text{i}\left|\bm{\b}_{\mu}\right|\text{Im}\(\bm{\b}_{\mu}\)}.  \label{btilde}
\eea  
When $\bm{\beta}_\mu=\im \beta_\mu$ is purely imaginary, we have $\bm{\beta}=\im \beta$ and $\bm{\tilde{\beta}}=\im \tilde{\beta}$, then the relation \eqref{btilde} is simplified to
\be \label{unb}
\tilde{\b}=\sqrt{\b_\mu^2+4\pi^2\rho_c}.
\ee 
This coordinate transformation \eqref{tildev} is uniquely fixed by requiring that it does not change the horizon area (black hole entropy) and the momentum which are supposed to be invariant under {the} \ttbar flow. Clearly, the undeformed CFT and the deformed CFT have the same spatial dimension. Note that the undeformed CFT and the seed CFT
are related by a Weyl transformation: 
\be 
dvd\bar{v}=\frac{(1-\rho_c L_0)(1-\rho_c \bar{L}_0)}{(1-\rho_c^2L_0\bar{L}_0)^2}d\tilde{v}d\bar{\tilde{v}}\equiv W^2 d\tilde{v}d\bar{\tilde{v}} \label{eq:weyl}, \quad W = \frac{|1+\frac{\rho_c\pi^2}{\bm{\beta}^2}|}{1-\frac{\rho_c^2\pi^4}{|\bm{\beta}|^4}},
\ee 
therefore, the undeformed energy and momentum are 
\bea 
\tilde{M}&=&\frac{c \tilde{\bm{\w}}}{12\pi}\frac{\(L_0+\bar{L}_0\)\(1+L_0\bar{L}_0\r_c^2\)-4\r_c L_0\bar{L}_0}{\(1-L_0\bar{L}_0\r_c^2\)},\nonumber \\
&=&-\frac{c\pi \bm{\w}_{\mu}}{3}\frac{(\bm{\b}_{\mu}+\bm{\bar{\b}}_{\mu})^2(\left|\bm{\b}_{\mu}\right|^2+2\pi^2\r_c)-2\left|\bm{\b}_{\mu}\right|^4}{\left|\bm{\b}_{\mu}\right|^3\sqrt{\left|\bm{\b}_{\mu}\right|^2+4\pi^2\r_c} \left(\left|\bm{\b}_{\mu}\right|+\sqrt{\left|\bm{\b}_{\mu}\right|^2+4\pi^2\r_c}\right)^2},\\
\tilde{J}&=& J_\mu, 
\eea 
and  they satisfy
\bea 
M_\mu = \frac{2\tilde{\bm{\w}}}{\mu}\left[1-\sqrt{1-\frac{\mu \tilde{M}}{\tilde{\bm{\w}}}-\frac{\mu^2\tilde{J}^2}{4\tilde{\bm{\w}}^2}}\;\right],
\eea 
as expected from the field theory result.
Substituting \eqref{bhl0} into \eqref{deformedonshell} and using the relation \eqref{bmu}, we find that the deformed on-shell action of the BTZ black hole is
\bea  \label{eq:onshell2}
I_E^{[\mu]}&=&-\frac{c\pi A_\mu}{3}\frac{1}{|\bm{\b_}\mu|^2+|\bm{\b}_\mu|\sqrt{4\pi^2\rho_c+|\bm{\b}_\mu|^2}}\\
&=&-\frac{cA_\mu}{6\pi}\frac{\pi^2}{|\bm{\beta}_\mu|^2}{+}\rho_c\frac{cA_\mu}{6\pi}\frac{\pi^4}{|\bm{\beta}_\mu|^2}+\mathcal{O}(\rho_c^2) \nonumber\\
&=&I_E^{(0)}+\mu \int d^2 x \sqrt{\gamma}\, \text{\ttbar}+\mathcal{O}(\mu^2), \label{bhexpand}
\eea
and the expansion \eqref{bhexpand} coincides exactly with the field theory result \eqref{fieldaction}, where $A_\mu=|\bm{\w}_\mu|^2\text{Im}(\bm{\tau}_\mu)$ is the area of the deformed torus. 
\subsection{Thermodynamics}
After the \ttbar deformation, the entropy of the black hole is not deformed, thus it is equal to\footnote{which is also equal to the black hole entropy of the undeformed theory.}
\bea \label{dBH}
S_{BH}^{[\mu]}=\frac{c\pi }{3}\frac{A_0}{|\bm{\beta}|^2}=\frac{c\pi}{3}\frac{A_\mu}{|\bm{\b}_\mu|^2}\frac{1}{\sqrt{1+\frac{4\pi^2\rho_c}{|\bm{\beta}_\mu|^2}}}=\frac{c\pi}{3}\frac{\text{Im}(\bm{\tau}_\mu)}{|\bm{\tau}_\mu|^2\sqrt{1+\frac{4\pi^2}{|\bm{\tau}_\mu|^2}\frac{\rho_c}{|\bm{\w}_\mu|^2}}},
\eea 
which not only depends on the modular parameter but also depends on the size of the system. For example, the deformation is insignificant when the torus is infinitely long. This is because that the effective deformation parameter is the dimensionless quantity $\rho_c/|\bm{\w}_\mu|^2$.
As expected, the deformed on-shell action still equals to the free energy
\bea 
I_{E}^{[\mu]}=\text{Im}(\bm{\beta}_\mu)M_\mu+\text{Re}(\bm{\beta}_\mu)J_\mu-S_{BH}^{[\mu]},
\eea 
even though both sides of this equation are deformed in a non-trivial way.
To examine the first law of thermodynamics, we treat $M_\mu$, $J_\mu$ and $S_{BH}^{[\mu]}$ as functions of $\bm{\beta}_\mu$ and $\bm{\bar{\beta}}_\mu$ and do the variation with respect to them. It turns out the first law is in the same form of the undeformed one
\be 
\text{Im}(\bm{\beta}_\mu)\delta M_\mu+\text{Re}(\bm{\beta}_\mu)\delta J_\mu=\delta S_{BH}^{[\mu]},
\ee 
but the Smarr relation takes a different form
\bea 
\text{Im}(\bm{\beta}_\mu)M_\mu+\text{Re}(\bm{\beta}_\mu)J_\mu=S_{BH}^{[\mu]}\frac{1}{1+\sqrt{1+\frac{4\pi^2\rho_c}{|\bm{\beta}_\mu|^2}}}.
\eea 
Interestingly, the Smarr relation looks like the standard one if we rewrite $\bm{\beta}_\mu$ in terms of $\bm{\beta}$ via \eqref{bmu}
\be 
\text{Im}(\bm{\beta})M_\mu+\text{Re}(\bm{\beta})J_\mu=\frac{1}{2}S_{BH}^{[\mu]}.
\ee 
On the other hand, the Cardy formula is now 
\bea 
S_{BH}^{[\mu]}=\sqrt{\frac{c\pi\bm{\w}_\mu}{6}{(M_\mu+\im J_\mu)}{}-\frac{c\pi\mu}{24}(M_\mu^2+J_\mu^2)}+\sqrt{\frac{c\pi\bm{\bar{\w}_\mu}}{6}{(M_\mu-\im J_\mu)}{}-\frac{c\pi\mu}{24}(M_\mu^2+J_\mu^2)}.\nonumber\\
\eea 
Moreover, the entropy of the black hole can also be found by using the Bekenstein-Hawking formula $S_{BH}^{[\mu]}=A_{\text{horizon}}/4G_N$. As a crosscheck of the validity of the deformed on-shell action, we can also derive the black hole entropy from the \Renyi entropies of the entire system using the replica trick:
\bea 
S_{BH}^{[\mu]}&=&{\lim_{n \rightarrow 1}}\frac{1}{1-n}\log \frac{Z_n}{Z_1^n}= {\lim_{n \rightarrow 1}}\frac{1}{1-n}\log \frac{e^{-I^{[\mu]}_{\text{on-shell,BTZ}}(n\bm{\tau}_\mu)}}{e^{-n I^{[\mu]}_{\text{on-shell,BTZ}}(\bm{\tau}_\mu)}}\nonumber \\
&=&{\lim_{n \rightarrow 1}} \frac{cn\,\text{Im}(\bm{\tau}_\mu)|\bm{\omega}_\mu^2|\(\sqrt{1+\frac{4\pi^2\rho_c}{|\bm{\omega}_\mu \bm{\tau}_\mu|^2}}-\sqrt{1+\frac{4\pi^2\rho_c}{|\bm{\omega}_\mu \bm{\tau}_\mu|^2n^2}}\)}{12(n-1)\pi\rho_c}\nonumber\\
&=&\frac{c\pi}{3}\frac{\text{Im}(\bm{\tau}_\mu)}{|\bm{\tau}_\mu|^2\sqrt{1+\frac{4\pi^2}{|\bm{\tau}_\mu|^2}\frac{\rho_c}{|\bm{\w}_\mu|^2}}},
\eea 
which is the same as \eqref{dBH}.
\subsection{Hawking-Page transition and Modular transformation}
\label{section:modular}

In AdS$_3$ gravity theory, the Hawking-Page phase transition happens when the free energy of the global AdS equals that of the BTZ black hole and there is only one transition point which is at $|\bm{\tau}|=1$. The free energy of the deformed BTZ and the deformed global AdS are \eqref{eq:onshell1} and \eqref{eq:onshell2}, which we respectively rewrite here in terms of $(\bm{\w}_\mu,\bm{\beta}_\mu)$: 
\bea 
&&I_{E,\text{BTZ}}^{[\mu]}=-\frac{c\pi}{3}\frac{\text{Im}(\bm{\tau}_\mu)}{|\bm{\tau}_\mu|^2}\frac{1}{1+\sqrt{1+\frac{4\pi^2\rho_c}{|\bm{\w_}\mu\bm{\tau}_\mu|^2 }}},\nonumber \\
&&I_{E,\text{global}}^{[\mu]}=-\frac{c\pi}{3}\frac{\text{Im}(\bm{\tau}_\mu)}{1+\sqrt{1+\frac{4\pi^2\rho_c}{|\bm{\w}_\mu|^2 }}}\nonumber.
\eea 
By equating them, we find that the Hawking-Page phase transition point remains unchanged at $|\bm{\tau}_\mu|=1$.
Clearly, the deformed on-shell actions $I_{E,\text{global}}^{[\mu]}$ and $I_{E,\text{BTZ}}^{[\mu]}$ are interrelated through the following modular transformation
\be 
\bm{\tau}_\mu\leftrightarrow -\frac{1}{\bm{\tau}_\mu},\quad \rho_c \leftrightarrow \frac{\rho_c}{|\bm{\tau}_\mu|^2}.
\ee 
This transformation aligns with the modular property exhibited by the \ttbar deformed torus partition function, as proved in \cite{Datta:2018thy,Apolo:2023aho}. To further support our argument regarding modular transformation, let us proceed to compute the deformed on-shell action for other $SL(2,\mathbb{Z})$ black holes. It is noteworthy that since the \ttbar deformation alters the temperature, the modular transformation does not commute with the \ttbar deformation. In other words, we have
 \be 
 \bm{\tau}_{\mu}(\bm{\tau})\neq -\frac{1}{\bm{\tau}_\mu(-\frac{1}{\bm{\tau}})}.
 \ee 
Therefore, it is not convenient to first construct the $SL(2,\mathbb{Z})$ black holes and then apply the \ttbar deformation. Instead, we will directly construct the \ttbar deformed $SL(2,\mathbb{Z})$ black holes following the approach outlined in \cite{Dijkgraaf:2000fq}. Starting from the deformed global AdS metric \eqref{gdemetric} with the following two complex periods\footnote{Here, we introduce the subscript $m$ to denote the modular transformation and distinguish $c_m$ from the CFT central charge $c$.}
\be 
\bm{\w}'_\mu=\bm{\w}_\mu,\quad \bm{\beta}'_\mu=\bm{\w}_\mu \frac{a_m \bm{\tau}_\mu+b_m}{c_m \bm{\tau}_\mu+d_m},\quad 
 \begin{pmatrix}
 	a_m& b_m\\
 	c_m&d_m
 \end{pmatrix}\in SL(2,\mathbb{Z}),\quad c_m\geq 0,
 \ee 
and making the following transformations
\be 
 V= \frac{1}{c_m\bm{\tau}_\mu+d_m}\hat{V}, \quad z=\frac{\hat{z}}{|c_m\bm{\tau}_\mu+d_m|},\quad \rho_c= \frac{\hat{\rho_c}}{|c_m\bm{\tau}_\mu+d_m|^2}\label{tran},
\ee
we end up with the deformed $SL(2,\mathbb{Z})$ black hole metric   
which is in the form of \eqref{demetric}
with
\be 
\hat{L}_0=-\frac{\pi^2}{\bm{\omega}_\mu^2(c_m\bm{\tau}_\mu+d_m)^2},\quad \bar{\hat{L}}_0=-\frac{\pi^2}{\bm{\bar{\omega}}_\mu^2(c_m\bm{\bar{\tau}}_\mu+d_m)^2}.
\ee 
The two complex periods of the deformed $SL(2,{\mathbb{Z}})$ black hole are given by
\be 
\bm{\hat{\w}}_\mu=\bm{\omega}_\mu (a_m \bm{\tau}_\mu+b_m),\quad \bm{\hat{\beta}}_\mu=\bm{\omega}_\mu (c_m \bm{\tau}_\mu+d_m),
\ee 
as expected for the $SL(2,{\mathbb{Z}})$ black hole. Since the coordinate transformations \eqref{tran} will not change the on-shell action, we conclude that
\bea 
I_{E,SL(2,{Z})}^{[\mu]}=I_{E,\text{global}}^{[\mu]}\left[\frac{a_m \bm{\tau}_\mu+b_m}{c_m \bm{\tau}_\mu+d_m}\Big| \frac{\hat{\rho}_c}{|c_m\bm{\tau}_\mu+d_m|^2}\right]=-\frac{c\pi}{3}\frac{\text{Im}(\frac{a_m \bm{\tau}_\mu+b_m}{c_m \bm{\tau}_\mu+d_m})}{1+\sqrt{1+\frac{4\pi^2\hat{\rho}_c}{|\bm{\w}_\mu|^2|c_m\bm{\tau}_\mu+d_m|^2 }}},
\eea 
which is exactly the modular transformation found in \cite{Datta:2018thy}. From the holographic perspective, the modular transformation can be readily understood through the transformations given in equation \eqref{tran}. The deformation is irrelevant, and as a result, the deformation parameter $\mu$ has a dimension and scales with the radial coordinate $z$ under a Weyl transformation. 
\\Analogous to the undeformed theory, a modular-invariant semi-classical partition function can be constructed by summing over all bulk saddle solutions \cite{Yin:2007gv,Witten:2007kt,Manschot:2007zb,Maloney:2007ud}: 
\bea 
Z^{[\mu]}_{\text{torus}}=\sum_{\text{saddles}}e^{-I^{[\mu]}_{E,\text{saddles}}},
\eea 
as the modular transformation simply maps a saddle to another one. However, this semi-classical partition function is not holomorphic factorized because $\tau_\mu$ and $\bar{\tau}_\mu$ transform collectively. To further illustrate that the modular property persists even when the holomorphic and anti-holomorphic sectors are asymmetric, we will explore the \ttbar deformation of topologically massive gravity, which is dual to a conformal field theory  with distinct holomorphic and anti-holomorphic central charges, i.e. $c_L\neq c_R$\cite{Alvarez-Gaume:1983ihn,Kraus:2005zm}. The AdS/CFT correspondence for TMG is elegantly established in \cite{Skenderis:2009nt}. For the calculation of the deformed on-shell action, we solely require the expression of the holographic energy-stress tensor to define the \ttbar deformation, as we shall demonstrate subsequently.

\section{Topologically massive gravity theory}
\renewcommand{\theequation}{5.\arabic{equation}}
\setcounter{equation}{0}
In this section, we consider the Euclidean version of the TMG theory whose action is given by
\bea 
I_{\text{TMG}}&=&-\frac{1}{16\pi G_N}\int d^3x(\mathcal{L}_{EH}+\mathcal{L}_{CS}),\\
\mathcal{L}_{CS}&=&\frac{ \a }{2 } \sqrt{g}\epsilon^{\lambda\mu\nu}\Gamma^\rho_{\lambda\sigma}\(\partial_\mu \Gamma^\sigma_{\rho\nu}+\frac{2}{3}\Gamma^\sigma_{\mu\tau}\Gamma^\tau_{\nu\rho}\),
\eea 
where $\mathcal{L}_{CS}$ is the gravitational Chern-Simons term. Notably, we assume the parameter $\alpha$ to be real, ensuring that the deformed metric remains real\footnote{For the Euclidean TMG, derived through a Wick rotation from the Lorentzian theory, the parameter $\alpha$ is purely imaginary because $\mathcal{L}_{CS}$ has odd parity.}.
The equation of motion for this theory is as follows
\be 
R_{\mu\nu}-\frac{1}{2}g_{\mu\nu}R-g_{\mu\nu}+ {\a } C_{\mu\nu}=0,
\ee 
where $C_{\mu\nu}$ is the Cotton tensor, defined as
\be
C_{\mu\nu}=\epsilon_\mu^{\alpha\beta}\nabla_\alpha(R_{\beta\nu}-\frac{1}{4}g_{\beta\nu}R).
\ee 
Consequently, the Einstein solutions \eqref{FG} also satisfy the equations of the TMG theory. However, the holographic stress tensor is modified to \cite{Kraus:2005zm,Solodukhin:2005ah}
\bea \label{eq:tmgstress}
T_{ij}=\frac{1}{8\pi G_N}\(g_{ij}^{(2)}-g_{ij}^{(0)}g_{kl}^{(2)}g^{kl}_{(0)}\)+\frac{ \alpha}{16\pi G }g^{kl}_{(0)}\(\epsilon_{lj}g_{ik}^{(2)}+\epsilon_{li}g_{jk}^{(2)}\),
\eea 
where $\epsilon_{ij}$ is the two-dimensional Levi-Civita tensor. Furthermore, the actions of global AdS and BTZ black hole in TMG are given in \cite{Kraus:2005vz,Kraus:2006nb}:
\be\label{On-shell Action TMG}
I_{\text{TMG},\text{global}}^{[0]}=\frac{\im \pi k}{2}\bm{\tau}-\frac{\im \pi \bar{k}}{2}\bm{\bar{\tau}},\quad I_{\text{TMG},\text{BTZ}}^{[0]}=-\frac{\im \pi k}{2\bm{\tau}}+\frac{\im \pi \bar{k}}{2\bm{\bar{\tau}}},
\ee
where the involved parameters $k$ and $\bar{k}$ are determined by the relations 
\be 
k-\bar{k}=\frac{\im\alpha}{2 G_N},\quad k+\bar{k}=\frac{1}{2G_N}.
\ee
In the Appendix \ref{TMG Action}, we give a simple derivation\footnote{The computation of the on-shell action is a little subtle. For example, one cannot obtain the consistent results by substituting the BTZ solution \eqref{eq:sbtz} into the Euclidean action of the TMG theory. In \cite{Kraus:2005vz,Kraus:2006nb}, the on-shell actions were derived in an indirect way. } of them based on the Smarr relation.

\subsection{\ttbar deformation}
The \ttbar deformation operator is still defined by
\bea 
\mathcal{O}_{\text{\ttbar}}=\frac{1}{8}\(T_{\alpha\beta}T^{\alpha\beta}-(T^\alpha_\alpha)^2\)=T_{vv}T_{\bar{v}\bar{v}}-T_{v\bar{v}}^2,
\eea 
therefore, the flow equation is still solved by \eqref{metricflowm} and \eqref{energyflowm}. Substituting the stress tensor \eqref{eq:tmgstress} into \eqref{metricflowm} leads to
\bea \label{tmggm}
\gamma_{\alpha\beta}^{[\mu]}dx^\alpha dx^\beta=\left(dv+\frac{\mu \bar{k}}{4\pi}\bar{L}_0 d\bar{v} \right) \left(d\bar{v}+\frac{\mu k}{4\pi}L_0dv \right)\equiv dVd\bar{V},
\eea 
which implies that
\be \label{dic}
V=v+\frac{\mu \bar{k}}{4\pi}\bar{L}_0 \bar{v},\quad \bar{V}=\bar{v}+\frac{\mu k}{4\pi}L_0 v,
\ee 
and 
\be 
\bm{\omega}_\mu=\bm{\omega}+\frac{\mu \bar{k}}{4\pi}\bar{L}_0\bar{\bm{\omega}},\quad \bm{\beta}_\mu=\bm{\beta}+\frac{\mu \bar{k}}{4\pi}\bar{L}_0\bar{\bm{\beta}}.
\ee 
Notice that \eqref{tmggm} can not be interpreted as an induced metric on a cut-off surface, hence, the cut-off (or glue-on) proposal for \ttbar deformation is no longer valid. The deformed on-shell action consists of two parts \eqref{onshell} which are convenient to express them in terms of the quantities in the seed theory:
\bea 
I_{\text{TMG}}^{[\mu]}=I_{E}^{[0]}-\mu A_0\frac{k\bar{k}}{4\pi^2}L_0\bar{L}_0,
\eea 
where $A_0$ is the area of the seed torus. In the end, we will use the dictionary \eqref{dic} to rewrite it in terms of the quantities in the deformed theory.
Until now, our discussion is very general. It is enough to demonstrate the modular transformation between the deformed global AdS and deformed BTZ and the other case can be generated trivially as we did in section \ref{section:modular}.
\\In this case, the global AdS solution corresponds to
\bea 
L_0=\bar{L}_0=-\frac{\pi^2}{\bm{\w}^2},
\eea 
and we find that the deformed on-shell action is
\bea  
I_{\text{TMG},\text{global}}^{[\mu]}&=&\frac{\im \pi}{2}(k\bm{\tau}-\bar{k}\bm{\bar{\tau}})+\frac{\im}{2}\mu(\bm{\tau}-\bm{\bar{\tau}}) \frac{|k|^2\pi^2}{4|\bm{\omega}|^2},\\
&=&{\im \pi} k\frac{\bm{\tau}_\mu}{1-\frac{(k-\bar{k})\pi \mu}{4|\bm{\w}_\mu|^2}+\sqrt{\frac{k\pi\mu}{|\bm{\w_}\mu|^2}+(1-\frac{(k-\bar{k})\pi\mu)}{4|\bm{\w}_\mu|^2})^2}}\nonumber \\
&-&\im \pi \bar{k}\frac{\bm{\bar{\tau}}_\mu}{1+\frac{(k-\bar{k})\pi \mu}{4|\bm{\w}_\mu|^2}+\sqrt{\frac{\bar{k}\pi\mu}{|\bm{\w}_\mu|^2}+(1+\frac{(k-\bar{k})\pi\mu)}{4|\bm{\w}_\mu|^2})^2}}.
\eea  
Moreover, the BTZ solution corresponds to
\bea 
L_0=-\frac{\pi^2}{\bm{\omega}^2\bm{\tau}^2},\quad \bar{L}_0=-\frac{\pi^2}{\bm{\bar{\omega}}^2\bm{\bar{\tau}}^2}.
\eea 
Then, we find that the deformed on-shell action is 
\bea 
I_{\text{TMG},\text{BTZ}}^{[\mu]}&=&-\frac{\im \pi k}{2\bm{\tau}}+\frac{\im \pi \bar{k}}{2\bm{\bar{\tau}}}+\frac{\im}{2}\mu (\bm{\tau}-\bm{\bar{\tau}})\frac{|k|^2\pi^2}{4|\bm{\omega}|^2|\bm{\tau}|^4},\\
&=&-\frac{\im k \pi}{\bm{\tau}_\mu}\frac{1}{1-\frac{(k-\bar{k})\pi \mu}{4|\bm{\w}_\mu|^2|\bm{\tau}_\mu|^2}+\sqrt{\frac{k\pi\mu}{|\bm{\w}_\mu|^2|\bm{\tau}_\mu|^2}+(1-\frac{(k-\bar{k})\pi\mu)}{4|\bm{\w}_\mu|^2|\bm{\tau}_\mu|^2})^2}}\nonumber \\
&+&\frac{\im \bar{k}\pi}{\bm{\bar{\tau}}_\mu}\frac{1}{1+\frac{(k-\bar{k})\pi \mu}{4|\bm{\w}_\mu|^2|\bm{\tau}_\mu|^2}+\sqrt{\frac{\bar{k}\pi\mu}{|\bm{\w}_\mu|^2|\bm{\tau}_\mu|^2}+(1+\frac{(k-\bar{k})\pi\mu)}{4|\bm{\w}_\mu|^2|\bm{\tau}_\mu|^2})^2}}.
\eea 
Clearly, these two deformed on-shell actions are related by the same modular transformation:
\bea 
\bm{\tau}_\mu \leftrightarrow -\frac{1}{\bm{\tau}_\mu},\quad \mu\leftrightarrow \frac{\mu}{|\bm{\tau}_\mu|^2}.
\eea 
It has been demonstrated in Section \ref{section:modular}, the change of the deformation parameter under the modular transformation can be understood from the fact that the \ttbar deformation is irrelevant. We proceed to investigate the modular properties of the root-\ttbar deformation, which is classically marginal. It is worth noting that the root-\ttbar deformation is both compatible with the \ttbar deformation and also preserves classical integrability \cite{Borsato:2022tmu}. Although the quantum aspects of root-\ttbar deformation are less understood\footnote{See \cite{Ebert:2024zwv} for some recent progress.}, a holographic description has been proposed in \cite{Ebert:2023tih}, which bears similarity to the mixed-boundary-condition description of the \ttbar deformation. In particular, the spectrum of the deformed theory is accurately reproduced. The authors of \cite{Ebert:2023tih} also suggested that the root-\ttbar deformed CFT might possess certain modular properties. In particular, in \cite{Tempo:2022ndz} it was shown that S-modular transformation is preserved. We provide some evidences by evaluating the deformed on-shell action, as we did for the \ttbar deformation.

\section{Root-\ttbar deformation}
\renewcommand{\theequation}{6.\arabic{equation}}
\setcounter{equation}{0}
Similar to the \ttbar deformation, Root-\ttbar deformation is also defined by a flow equation
\bea 
\frac{d S^{(\mu)}}{d\mu}=\int d^2 x\sqrt{\gamma}\mathcal{R},\quad \mathcal{R}=\sqrt{\frac{1}{2}T^{\alpha\beta}T_{\alpha\beta}-\frac{1}{4}(T^\alpha_\alpha)^2}.
\eea 
 Because $\mathcal{R}$ is a marginal multi-trace operator, there is no need to add a boundary term to the on-shell action like the one in \eqref{onshell}. The solution of the flow equation is not unique but it was argued in \cite{Ebert:2023tih}, the proper root-\ttbar deformed boundary metric and stress tensor are given by
\be 
\gamma_{\alpha\beta}^{(\mu)}=\cosh(\mu)\gamma_{\alpha\beta}^{(0)}+\frac{\sinh(\mu)}{\mathcal{R}^{(0)}}\tilde{T}_{\alpha\beta}^{(0)},\quad \tilde{T}^{(\mu)}_{\alpha\beta}=\cosh(\mu)\tilde{T}_{\alpha\beta}^{(0)}+\sinh(\mu)\mathcal{R}^{(0)}\gamma_{\alpha\beta}^{(0)},
\ee 
where $\tilde{T}_{\alpha\beta}$ is the traceless part of the stress tensor $\tilde{T}_{\alpha\beta}=T_{\alpha\beta}-\frac{1}{2}T\gamma_{\alpha\beta}$. We will again start from the seed solution \eqref{FG}, for which the stress tensor is $8\pi G_NT_{\alpha\beta}=\text{diag}(L_0,\bar{L}_0),\,\tilde{T}_{\alpha\beta}=T_{\alpha\beta}$ and the operator $\mathcal{R}^{(0)}$ equals
\be 
\mathcal{R}^{(0)}=\frac{\sqrt{L_0\bar{L}_0}}{4\pi G_N}.
\ee 
Then it is easy to obtain the deformed boundary metric
\be
 \gamma_{\alpha\beta}^{(\mu)}=\cosh(\mu)g_{\alpha\beta}^{(0)}+\frac{\sinh(\mu)}{2\sqrt{L_0\bar{L}_0}}g_{\alpha\beta}^{(2)},
\ee 
and the deformed stress tensor
\be 
\tilde{T}^{(\mu)}_{\alpha\beta}=\frac{1}{8\pi G_N }\(\cosh(\mu)g_{\alpha\beta}^{(2)}+2\sqrt{L_0\bar{L}_0}\sinh(\mu)g_{\alpha\beta}^{(0)}\),\quad T^{(\mu)}=T^{(0)},
\ee 
where we have used the fact that for a classically marginal deformation, the trace of the stress tensor does not flow.
Again we can introduce the flat coordinates on the deformed boundary and write the deformed boundary metric as follows
\be 
\gamma_{\alpha\beta}^{(\mu)}dx^\alpha dx^\beta=dVd\bar{V},
\ee 
where 
\bea  
&&dV=\cosh\(\frac{\mu}{2}\)dv+\sqrt{\frac{\bar{L}_0}{L_0}}\sinh \(\frac{\mu}{2} \)d\bar{v},\quad d\bar{V}=\cosh \(\frac{\mu}{2} \)d\bar{v}+\sqrt{\frac{L_0}{\bar{L}_0}}\sinh \(\frac{\mu}{2} \)dv,\nonumber \\
&&dv=\cosh \(\frac{\mu}{2} \)dV-\sqrt{\frac{\bar{L}_0}{L_0}}\sinh \(\frac{\mu}{2} \)d\bar{V},\quad d\bar{v}=\cosh \(\frac{\mu}{2} \)d\bar{V}-\sqrt{\frac{L_0}{\bar{L}_0}}\sinh \(\frac{\mu}{2} \)dV.\nonumber 
\eea  
In global AdS$_3$ solution, $L_0=\bar{L}_0=-\frac{\pi^2}{\bm{\omega}^2}$ then the two periods of the deformed torus are
\be 
\bm{\omega}_\mu=\(\cosh \(\frac{\mu}{2} \)-\sinh \(\frac{\mu}{2} \)\)\bm{\omega}=e^{-\frac{\mu}{2}}\bm{\omega},\quad \bm{\beta}_\mu=\cosh\(\frac{\mu}{2} \)\bm{\beta}+\sinh \(\frac{\mu}{2} \)\bm{\bar{{\beta}}}.
\ee 
In this case, the deformed on-shell action is given by the first term of \eqref{deformedonshell}:
\be 
I^{[\mu]}_{E,global}=-\frac{c}{6\pi}\bm{\omega}_\mu^2\text{Im}(\bm{\tau}_\mu)\frac{\pi^2}{\bm{\omega}_\mu^2 e^\mu}=-e^{-\mu}\frac{c\pi}{6}\text{Im}(\bm{\tau}_\mu).
\ee 
On the other hand, in the BTZ solution with two complex periods $\bm{\omega}=\omega_1+\im \omega_2,\bm{\beta}=\beta_1+\im \beta_2$, we have
\be 
L_0=-\frac{\pi^2}{\bm{\beta}^2},\quad \bar{L}_0=-\frac{\pi^2}{\bm{\bar{\beta}^2}}.
\ee 
Then the deformed periods are 
\bea  
&& {\bm{\omega}_\mu= \bm{\omega} \cosh \( \frac{\mu}{2} \) - \( \frac{\bm{\beta}\bm{\bar{\omega}}}{\bm{\bar{\beta}}} \) \sinh \( \frac{\mu}{2} \) ,} \\
&&\bm{\beta}_\mu=e^{-\frac{\mu}{2}}\bm{\beta}.
\eea 
We find that the deformed on-shell action is
\be 
I^{[\mu]}_{\text{E,BTZ}}=-\frac{c}{6\pi}\bm{\omega}_\mu^2\text{Im}(\bm{\tau}_\mu)\frac{\pi^2}{|\bm{\beta_\mu}|^2 e^{\mu}}=e^{-\mu}\frac{c\pi}{6}\frac{\text{Im}(\bm{\tau}_\mu)}{|\bm{\tau}_\mu|^2},
\ee 
which is related to $I^{[\mu]}_{\text{E,global}}$ simply by $\bm{\tau}_\mu \leftrightarrow -\frac{1}{\bm{\tau}_\mu}$. Notice that in this case, the deformation parameter $\mu$ is not changed under the modular transformation. Therefore, our on-shell action calculation suggests that the modular transformation of the root-\ttbar deformed CFT is the same as that of the undeformed one.

\section{Discussion}
In this work, we study the modular properties of the (root-) \ttbar deformed torus partition function of a 2d CFT (with a gravitational anomaly) from the holographic perspective by computing the on-shell actions of various saddle solutions of the dual gravity theories. 
\\For a 2d \ttbar deformed CFT, we reproduce the modular transformation \eqref{moz} of the partition function. From the bulk point of view, the transformation rule of the deformation parameter $\mu$ arises because its dimension matches that of  the radial coordinate. 
We also find that the modular property \eqref{moz} is preserved even when the CFT has a gravitational anomaly and is dual to a TMG theory. Moreover, we show that the root-\ttbar deformation does not alter the behavior of the partition function under the modular transformation, and hence it is the same as that of the undeformed CFT. 
It should be pointed out that deriving these properties from the field theory side using the same strategy as the one applied in \cite{Datta:2018thy} would be very interesting. 
\\Another closely related possible future direction is to study the modular property of the $\text{J}\bar{\text{T}}$ deformed partition function holographically in the mixed-boundary-condition proposal \cite{Bzowski:2018pcy}. The $\text{J}\bar{\text{T}}$ deformed partition function is covariant under modular transformation as shown in \cite{Aharony:2018ics}, in particular the deformation parameter also transforms. Therefore, we expect that it can also be easily understood from the bulk point of view.
\\Here we only focus on the classical contribution to the partition function. It is crucial but challenging to include the quantum corrections. For pure AdS gravity theory, the quantum correction{s arise} from summing over states corresponding to the Virasoro descendants. In a recent work \cite{He:2024pbp}, this 1-loop effect was studied by assuming that the only effect of the \ttbar deformation is to alter the temperature, and then the 1-loop contribution can be computed using the usual heat kernel method on a BTZ geometry. However, as shown in \cite{Guica:2019nzm}, the asymptotic symmetry algebra of the deformed theory is not a naive Virasoro algebra, but actually a state-dependent one. Thus, this requires further study from both the field theory and gravity theory perspectives.
\\In the end, we want to emphasize that the boundary term \eqref{bdy} in the deformed on-shell action and the holographic dictionary \eqref{dictionary} deserve further exploration. For example, as suggested in \cite{Jia:2023ttbar}, it may modify the holographic entanglement entropy. Furthermore, in an oncoming work \cite{newbanana}, we also find that naively applying the dictionary to the \ttbar deformed cone solution or banana solution \cite{Abajian:2023jye} will lead to wrong results.

\section*{Acknowledgments}
We thank Miao He, Yan Liu, Yang Lei and Huajia Wang for the valuable discussion. JT is supported by  the National Youth Fund No.12105289 and funds from the University of Chinese Academy of Sciences (UCAS), the Kavli Institute for Theoretical Sciences (KITS).
FO would like to thank Masahiro Nozaki very much for his supports. The work of FO was supported by funds from UCAS, KITS and The Chinese Academy of Sciences (CAS) President's International Fellowship Initiative (PI FI).

\appendix 
	
\section{Hubbard-Stratonovich method of \ttbar}
\label{append:ttbar}
\renewcommand{\theequation}{A.\arabic{equation}}
\setcounter{equation}{0}
In this appendix, we show some details  about the holographic interpretation of the \ttbar deformation mainly following \cite{Monica}.
Let $\mathcal{O}_A$ be a collection of local operators in CFT which are ``single trace" in the sense it is dual to some simple bulk field. For example, the energy-momentum tensor $T_{\alpha\beta}$ is treated as a single trace operator, because it is dual to the boundary metric $\gamma_{\alpha\beta}$. Then the double trace deformation can be schematically defined as 
\bea 
S=S_{0}+\frac{N^2\mu }{2}\int d^2x \sqrt{\gamma} \eta^{AB}\mathcal{O}_A\mathcal{O}_B,
\eea 
where $\mu$ is the deformation parameter, $\eta^{AB}$ are some constants for specifying a particular double trace operator and $\gamma_{\mu\nu}$ is the metric of the spacetime where the CFT is defined on. When we derive the holographic dual, we can assume that $\gamma_{\alpha\beta}$ is flat and restore it when it is necessary. The deformed generating function is defined as
\bea 
e^{-W^{(\mu)}[J^{(\mu)}]}=\int \mathcal{D}\psi \exp\(-S_0-N^2\int d^2x \(\frac{\mu}{2}\eta^{AB}\mathcal{O}_A\mathcal{O}_B+J^{(\mu)A}\mathcal{O}_A\)\).
\eea 
Following the Hubbard-Stratonovich method, we integrate in the fields $h_A$ by inserting the resolution of identity
\bea 
1=\mathcal{N}\int \mathcal{D}h\exp\(\frac{N^2}{2\mu}\int d^2 h^A(\eta^{-1})_{AB}h^B\),
\eea 
and combine the two quadratic terms which leads to
\bea 
e^{-W^{(\mu)}[J^{(\mu)}]} \!\!\! && \!\!\!\!\!=\mathcal{N}\int \mathcal{D}h\mathcal{D}\psi\exp\(-S_0+N^2\int d^2x \frac{1}{2\mu}(h^A+\mu \eta^{AC}\mathcal{O}_C)(\eta^{-1})_{AB}(h^B+\mu \eta^{BD}\mathcal{O}_D) \right. \nonumber \\
&& \left. -N^2\int d^2x  \mathcal{O}_A (J^{(\mu)A}+\mu \eta^{AB}\mathcal{O}_B+ h^A)                   \).
\eea 
Introducing the shifted source 
\bea 
\tilde{J}^A=J^{(\mu)A}+\mu \eta^{AB}\mathcal{O}_B+ h^A,
\eea 
then the generating function can be cast into
\bea
&&e^{-W^{(\mu)}[J^{(\mu)}]}=\mathcal{N}\int  \mathcal{D}h\, e^{-W^{(0)}[\tilde{J}]}\exp\(\frac{N^2}{2\mu} \int d^2 x (\tilde{J}^A-J^{(\mu)A})(\eta^{-1})_{AB}(\tilde{J}^B-J^{(\mu)B})\).
\;\;\;\;\;\;\;\;
\label{g2}
\eea 
Ultimately, we evaluate the $h$ integral with the saddle point approximation. Varying the exponent with respect to $h_A$ gives
\bea 
-\mathcal{O}_A+\frac{1}{\mu}(\eta^{-1})_{AB}(\tilde{J}^B-J^{(\mu)B})=0,
\eea 
with the solution
\bea 
\tilde{J}^A=\mu \eta^{AB}\mathcal{O}_B+J^{(\mu)A},
\eea 
Substituting it into \eqref{g2}, we end up with
\bea 
W^{(\mu)}[J^{(\mu)}]=W^{(0)}[J^{(\mu)A}+\mu \eta^{AB}\mathcal{O}_B]-\frac{N^2\mu }{2}\int d^2x \eta_{AB}\mathcal{O}_A\mathcal{O}_B.
\eea 
Note that this is only the result of the first-order deformation. To obtain the deformed theory on the whole \ttbar flow, we have to work out 1) the deformed source $J^{(\mu)}$; 2) the deformed functional $W^{(\mu)}$. The deformed source $J^{(\mu)}$ has been derived by the variation method that we review below.
Using the defining property of the generating function $\delta W=\frac{1}{2}\int d^2 x\sqrt{\gamma}T_{\alpha\beta}\delta\gamma^{\alpha\beta}$, we can obtain a flow equation 
\bea 
\frac{1}{2}\partial_\mu\(\int d^2x \sqrt{\gamma^{[\mu]}}T_{\alpha\beta}^{[\mu]}\delta \gamma_{\alpha\beta}^{[\mu]}\)=-\delta\(\int d^2x \sqrt{\gamma^{[\mu]}}\text{T}\bar{\text{T}}^{[\mu]}\),
\eea 
which can be solved by 
\bea 
&&\gamma_{\alpha\beta}^{[\mu]}=\gamma_{\alpha\beta}^{[0]}+\frac{1}{2}\mu \hat{T}_{\alpha\beta}^{[0]}+\frac{1}{16}\mu^2\hat{T}_{\alpha\rho}^{[0]}\hat{T}_{\sigma\beta}^{[0]}{\gamma^{[0]}}^{\rho\sigma},\label{metricflow}\\
&&\hat{T}^{[\mu]}_{\alpha\beta}=\hat{T}^{[0]}_{\alpha\beta}+\frac{1}{4}\mu\hat{T}_{\alpha\rho}^{[0]}\hat{T}_{\sigma\beta}^{[0]}{\gamma^{[0]}}^{\rho\sigma},\label{energyflow}
\eea 
where $\hat{T}_{\alpha\beta}=T_{\alpha\beta}-\gamma_{\alpha\beta}T$. Some details are the following. Explicitly, the flow equation is 
\bea 
\partial_{\mu}(\sqrt{\gamma} T_{\alpha\beta}\delta \gamma^{\alpha\beta})=-\frac{1}{4}\delta (\sqrt{\gamma}(T_{\alpha\beta}T^{\alpha\beta}-(T^\alpha_\alpha)^2))\equiv \delta(\sqrt{\gamma} \mathcal{O}_{\text{\ttbar}}).
\eea 
The left-hand-side is
\bea 
\partial_{\mu}(\sqrt{\gamma}T_{\alpha\beta})\delta \gamma^{\alpha\beta}+\sqrt{\gamma}T_{\alpha\beta}\delta(\partial_\mu \gamma^{\alpha\beta}),
\eea 
and the right-hand-side is
\bea 
&&-\frac{1}{4}\sqrt{\gamma}\(2(T_{\alpha\beta}\delta T^{\alpha\beta}+T^{\alpha\gamma}T^\beta_\gamma\delta\gamma_{\alpha\beta}-T\delta T)\)-\frac{1}{2}\sqrt{\gamma}\mathcal{O}_{\text{\ttbar}}\gamma_{\alpha\beta}\delta \gamma^{\alpha\beta}\\
&&=\sqrt{\gamma}\(C_1\delta\gamma^{\alpha\beta}+T_{\alpha\beta}\delta C_2\).
\eea 
Comparing the lhs and rhs, we get
\bea 
&&\partial_{\mu}(\sqrt{\gamma} T_{\alpha\beta})=C_1\sqrt{\gamma},\\
&&\partial_{\mu}\gamma^{\alpha\beta}=C_2,
\eea 
with
\bea 
&&C_1=-\frac{1}{2}\mathcal{O}_{\text{\ttbar}}\gamma_{\alpha\beta}+\frac{1}{2}T_{\alpha}^\delta T_{\delta \beta}-\frac{1}{2}TT_{\alpha\beta},\\
&&C_2=-\frac{1}{2}T^{\alpha\beta}+\frac{1}{2}T\gamma^{\alpha\beta}.
\eea 
The proposal is that $\gamma^{[\mu]}_{\alpha\beta}$ and $T^{[\mu]}_{\alpha\beta}$ are the sources and dual operators of the deformed holographic theory which is still an asymptotic AdS$_3$ gravity theory. The general asymptotic AdS$_3$ solution can be written in the Fefferman-Graham gauge as follows
\bea 
ds^2=g_{\alpha\beta}(\rho,x^\alpha)dx^\alpha dx^\beta+\frac{d\rho^2}{4\rho^2},\quad g_{\alpha\beta}(\rho,x^\alpha)=\frac{g_{\alpha\beta}^{(0)}}{\rho}+g_{\alpha\beta}^{(2)}+\rho g_{\alpha\beta}^{(4)},
\eea 
where $g^{(2)}$ corresponds to the initial expectation value of the CFT operator
\bea \label{energymap}
\hat{T}^{[0]}_{\alpha\beta}=\frac{1}{8\pi G_N}g_{\alpha\beta}^{(2)}.
\eea 
The solution \eqref{metricflow} implies that the deformed boundary metric is
\bea 
\gamma_{\alpha\beta}^{[\mu]}=g_{\alpha\beta}^{(0)}+\frac{\mu}{16\pi G_N}g_{\alpha\beta}^{(2)}+\frac{\mu^2}{(16\pi G_N)^2}g_{\alpha\beta}^{(4)}=\rho_c g_{\alpha\beta}(\rho_c),\quad \rho_c=\frac{\mu}{16\pi G_N }.
\eea 
To derive the deformed generating functional, we notice
\bea 
W^{(\mu)}&=&W^{(\mu-\delta\mu)}[J^{(\mu-\delta \mu)}+\delta \mu \eta^{AB}\mathcal{O}^{(\mu-\delta \mu)}]-\delta\mu \frac{N^2 }{2}\int d^2x \eta_{AB}[\mathcal{O}_A\mathcal{O}_B]^{(\mu)} \\
&=&W^{(0)}\left[J^{[\mu]}(J^{(0)})\right]-\frac{N^2}{2}\int_0^\mu dk \int dx^2 \eta_{AB}[\mathcal{O}_A\mathcal{O}_B]^{(k)}.
\eea 
In the case of \ttbar deformation, we have
\bea 
W^{(\mu)}&=&W^{(0)}[\gamma_{\alpha\beta}^{[\mu]}=g_{\alpha\beta}^{(0)}+\frac{\mu}{16\pi G_N}g_{\alpha\beta}^{(2)}+\frac{\mu^2}{(16\pi G_N)^2}g_{\alpha\beta}^{(4)}]-\int^\mu dk \int d^2 x\sqrt{\gamma^{(k)}}\text{\ttbar}^{(k)},\nonumber \\
&=&W^{(0)}[\gamma_{\alpha\beta}^{[\mu]}=g_{\alpha\beta}^{(0)}+\frac{\mu}{16\pi G_N}g_{\alpha\beta}^{(2)}+\frac{\mu^2}{(16\pi G_N)^2}g_{\alpha\beta}^{(4)}]-\mu \int d^2 x\sqrt{\gamma^{(0)}}\text{\ttbar}^{(0)},
\eea 
where  in the last step, we have used the fact that$ \sqrt{\gamma^{(k)}}\text{\ttbar}^{(k)}$ is a flow invariant and 
\bea 
\sqrt{\gamma^{(\mu)}}\text{\ttbar}^{(\mu)}=\sqrt{\gamma^{(0)}}\text{\ttbar}^{(0)}.
\eea 

\section{BTZ black holes with complex periods}
\renewcommand{\theequation}{B.\arabic{equation}}
\setcounter{equation}{0}
\label{generalEM}
In this appendix, we find the mass, angular momentum and entropy of a rotating BTZ black hole in terms of the complex periods of the torus on which the dual CFT lives. To do so, we first transform the BTZ metric \eqref{eq:sbtz} into the FG gauge \eqref{FG}. It is straightforward to verify that
\bea 
L_0&=&\frac{\(R_{+}+\text{i}R_{-}\)^2}{4}=\frac{\pi^2}{\(\b-\text{i}\theta\)^2}\equiv-\frac{\pi^2}{\bm{\b}^2},\\\nonumber
\bar{L}_0&=&\frac{\(R_{+}-\text{i}R_{-}\)^2}{4}=\frac{\pi^2}{\(\b+\text{i}\theta\)^2}\equiv-\frac{\pi^2}{\bar{\bm{\b}}^2}.
\eea
Then, from the holographic stress tensor \eqref{energymap}, one can derive the holographic mass and momentum
\bea 
M&=&-\int d\phi\, T_{\t\t}=\frac{c}{6}\(L_0+\bar{L}_0\)=\frac{\pi^2c}{3}\frac{\b^2-\theta^2}{\(\b^2+\theta^2\)^2},\\
J&=&-\int d\phi\, T_{\t\q}=-\frac{\text{i}c}{6}\(L_0-\bar{L}_0\)=\frac{\pi^2c}{3}\frac{2\b\theta}{\(\b^2+\theta^2\)^2}.
\eea 
Moreover, the entropy of the black hole is equal to
\be 
S_{BH}=\frac{2\pi R_{+}}{4G_N}=\frac{\pi c}{3}\(\sqrt{L_0}+\sqrt{\bar{L}_0}\)=\frac{\pi^2 c}{3}\frac{2\b}{\b^2+\theta^2}.
\ee 
Now to obtain a generic torus, we do a coordinate transformation
\be 
v \rightarrow \frac{\w}{2\pi}e^{\text{i}\a}v=\hat{v},
\ee  
or equivalently
\begin{eqnarray}
	\begin{pmatrix}
		\hat{\phi}\\
		\hat{t}
	\end{pmatrix}=\frac{\w}{2\pi}\begin{pmatrix}
		\cos\a & -\sin \a\\
		\sin\a & \cos\a
	\end{pmatrix}\begin{pmatrix}
		\phi \\
		t
	\end{pmatrix},
\end{eqnarray}
which implies that the periods are transformed as follows
\begin{eqnarray}
	\begin{pmatrix}
		\hat{\w}_1\\
		\hat{\w}_2
	\end{pmatrix}&=&\frac{\w}{2\pi}\begin{pmatrix}
		\cos\a & -\sin \a\\
		\sin\a & \cos\a
	\end{pmatrix}\begin{pmatrix}
		\w_1\\
		\w_2
	\end{pmatrix},\\
	\begin{pmatrix}
		\hat{\b}_1\\
		\hat{\b}_2
	\end{pmatrix}&=&\frac{\w}{2\pi}\begin{pmatrix}
		\cos\a & -\sin \a\\
		\sin\a & \cos\a
	\end{pmatrix}\begin{pmatrix}
		\b_1\\
		\b_2
	\end{pmatrix},
\end{eqnarray}
where $\w_1=2\pi, \w_2=0, \w_1+\im \w_2=\bm{\w},$ and $ \beta_1=\theta, \beta_2=\beta, \beta_1+\im \beta_2=\bm{\beta}$.
Noting that this transformation is a combination of Weyl transformation and rotation, the holographic mass and momentum should transform accordingly as 
\begin{eqnarray}
	\begin{pmatrix}
		\hat{J}\\
		\hat{M}
	\end{pmatrix}=\frac{2\pi}{\w}\begin{pmatrix}
		J\cos\a-M\sin\a\\
		J \sin\a+M\cos\a
	\end{pmatrix},
\end{eqnarray}
which leads to \eqref{eqm} and \eqref{eqj}. In addition, these two quantities correspond to the conserved charges of $\pa_{\hat{\f}}$  and $\pa_{\hat{t}}$, respectively.

\section{On-Shell action of TMG theory}\label{TMG Action}
\renewcommand{\theequation}{C.\arabic{equation}}
\setcounter{equation}{0}

In this appendix{,} we will derive the on-shell action \eqref{On-shell Action TMG} of TMG theory by assuming the Smarr relation still holds. The Smarr relation can be viewed as a relationship between the energy and  entropy and it holds separately for the holomorphic and anti-holomorphic sectors. Therefore, it is not modified even when $c_L\neq c_R$.
Starting from the following expressions 
\bea
T_{v v}=\frac{1+\text{i}\a}{8\pi G_N}L_0=\frac{k}{2\pi}L_0,\quad T_{\bar{v}\bar{v}}=\frac{1-\text{i}\a}{8\pi G_N}\bar{L}_0=\frac{\bar{k}}{2\pi}\bar{L}_0 ,
\eea
it is straightforward to obtain 
the holographic mass and momentum 
\bea
M&=&-\int d\f T_{t t}=\frac{\omega(k L_0+\bar{k}\bar{L}_0)}{2\pi},\\
J&=&-\int d\f T_{t\f}=\frac{-\text{i}\omega(k L_0-\bar{k}\bar{L}_0)}{2\pi}.
\eea
Here we have chosen a real $\omega$. Actually, by applying \eqref{eqm} and \eqref{eqj}, it is straightforward to obtain the result with complex period $\bm{\omega}$ as follows
\bea
M=\frac{\bm{\omega} k L_0+\bar{\bm{\omega}}\bar{k}\bar{L}_0}{2\pi}, \quad J=\frac{-\text{i}(\bm{\omega} k L_0-\bar{\bm{\omega}}\bar{k}\bar{L}_0)}{2\pi}.
\eea
Therefore, for global AdS the on-shell action is given by
\bea
I^{[0]}_{\text{TMG,global}}=\b_2 M+\b_1 J=\frac{-\text{i}(\bm{\b}\bm{\omega} k L_0-\bar{\bm{\b}}\bar{\bm{\omega}}\bar{k}\bar{L}_0)}{2\pi}.
\eea
Next, by applying $L_0=-\frac{\pi^2}{\bm{\omega}^2}$ and $\bm{\b}=\bm{\w}\bm{\tau}$, the above result can be simplified to 
\bea
I^{[0]}_{\text{TMG,global}}=\frac{\text{i}\pi k}{2}\bm{\tau}-\frac{\text{i}\pi \bar{k}}{2}\bar{\bm{\tau}}.
\eea
On the other hand, for the BTZ black hole, the on-shell action is as follows
\bea
I^{[0]}_{\text{TMG,BTZ}}=\b_2 M+\b_1 J-S_{BH}.
\eea
Using the Smarr relation
\bea
2(\b_2 M+\b_1 J)=S_{BH},
\eea
we find that the on-shell action equals to
\bea
I^{[0]}_{\text{TMG,BTZ}}=-(\b_2 M+\b_1 J)=-\frac{\text{i}\pi k}{2\bm{\tau}}+\frac{\text{i}\pi \bar{k}}{2\bar{\bm{\tau}}},
\eea
where we have used the fact $L_0=-\frac{\pi^2}{\bm{\b}^2}$ and $\bar{L}_0=-\frac{\pi^2}{\bar{\bm{\b}}^2}$.


\end{document}